\RequirePackage{fix-cm}
\documentclass[twocolumn,epjc3]{svjour3}  
\smartqed  
\RequirePackage{graphicx}
\RequirePackage{amsmath}
\journalname{Eur. Phys. J. C}
\begin{document}

\title{Phase control of multi-photon electron-positron pair creation from vacuum
}


\author{C.~K.~Li\thanksref{addr1,addr2}
        \and
        X.~X.~Zhou\thanksref{addr3}
        \and
B.~An\thanksref{addr4}
 \and
Y.~J.~Li\thanksref{addr4,addr5}
\and
N.~S.~Lin\thanksref{e1,addr5}
\and
Y.~Wan\thanksref{e2,addr1}
}

\thankstext{e1}{e-mail: phy.nslin@gmail.com}
\thankstext{e2}{e-mail: yangwan23@zzu.edu.cn}


\institute{Laboratory of Zhongyuan Light, School of Physics, Zhengzhou University, Zhengzhou 450001, China \label{addr1}
           \and
           Institute of Astronomy and Physics, Inner Mongolia University, Hohhot 010021, China \label{addr2}
           \and
           School of Management Science and Engineering, Anhui University of Finance and Economics, Bengbu 233030, China\label{addr3}
           \and
           State Key Laboratory for Tunnel Engineering, China University of Mining and Technology, Beijing 100083, China\label{addr4}
           \and
           School of Science, China University of Mining and Technology, Beijing 100083, China\label{addr5}
}

\date{Received: date / Accepted: date}

\maketitle

\begin{abstract}
We investigate the creation of electron-positron pairs by two spatiotemporally inhomogeneous electric fields with a relative phase, employing computational quantum field theory. We find that, when the two fields are closely spaced, the pair yield exhibits a cosine-like dependence on the relative phase. This suggests that the relative phase provides an effective way to enhance multi-photon transition channels. Furthermore, our analysis reveals that the response of pair-creation channels to the relative phase changes substantially with the photon order of the transition. For one-photon transitions, the rate exhibits a $2\pi$ periodicity, whereas a reduced periodicity of $\pi$ is observed for two-photon transitions. To clarify the underlying mechanism, we map the quantum field-theoretical framework onto a time-dependent perturbation approach. By extending this approach to $n$-photon processes, we show that the transition probability is periodic in the relative phase with period $2\pi/n$. This observation suggests that the relative phase offers an effective means of identifying the order of multi-photon transitions.
\end{abstract}

\section{Introduction}
\label{intro}
A celebrated non-perturbative prediction of quantum electrodynamics (QED) is the Sauter–Schwinger effect, the creation of electron–positron pairs from the vacuum driven by intense background fields \cite{Sauter,Euler,ADP1,AFed}.
This phenomenon has so far eluded experimental confirmation, since the peak intensity achievable with state-of-the-art laser facilities (currently on the order of $10^{23}~\text{W/cm}^2$) lies orders of magnitude below the critical intensity $I_{\mathrm{cr}} \sim 10^{29}~\text{W/cm}^2$. This critical intensity corresponds to the Schwinger critical field $E_{\mathrm{cr}} = m_e^2 c^3/(e\hbar) \approx 1.3 \times 10^{16}~\text{V/cm}$, where $m_e$ and $e$ denote the electron mass and charge, respectively \cite{Schwinger}.
Another significant mechanism for electron-positron pair creation is multi-photon pair creation, which has been observed using perturbative methods \cite{Brezin}. 
This phenomenon was detected at the Stanford Linear Accelerator Center (SLAC) in 1997 \cite{Burke}.
After the early pioneering multi-photon induced pair creation experiments, several worldwide laboratories, such as the Extreme-Light Infrastructure (ELI) \cite{ELI}, the eXawatt Center for Extreme Light Studies (XCELS) \cite{XCELS}, the European X-Ray Free-Electron Laser (EXFEL) \cite{XRFEL} and Station of Extreme Light (SEL) \cite{SEL}, are presently exploring new means to probe the quantum vacuum with very intense electromagnetic radiation fields. 
Notably, pair creation can already take place at laser intensities one to two orders of magnitude below the critical threshold. 
This enhanced sensitivity originates from the dynamically assisted Schwinger mechanism, where a strong low-frequency field and a weak high-frequency field act jointly to stimulate the creation process \cite{RS1,ADP2,IAA1,SVC1,HT1,YLR1,IAA2}.
Considerable attention has also been given to optimizing the spacetime structure of the external field in order to maximize the yield of created particles \cite{FH1,SSB1,CK1,FH2,SSD1,JU1}.

Recently, we have shown that, even when excluding photon exchange processes and Pauli blocking based on occupation numbers, a nontrivial mechanism remains through which electronic phases can significantly influence the dynamics of laser-induced electron-positron pair creation from vacuum \cite{CKL}.
In addition, previous investigations have demonstrated that the phase of external fields plays a significant role in pair creation, revealing novel features.
Two main lines of development may be distinguished.
On the one hand, the influence of the carrier-envelope phase (CEP) of laser pulses on pair creation processes has been extensively investigated \cite{NA1,MJAJ,AIT1,CB1,JJJ1}.
For example, the interference effects of different multi-photon pair-creation processes have been shown to depend sensitively on the CEP of the laser pulse \cite{MJAJ}.
It has been proposed that laser-induced pair creation could provide an alternative way to determine the CEP of ultra-intense laser pulses \cite{JJJ1}.
On the other hand, the pair creation can be affected by the relative phase between external fields \cite{KK1,SA1,ZLL2,CB2,JB2,CKL1,LNH1}. 
To analyze the dependence of the particle momentum distribution on this relative phase, the Phase-of-the-phase spectroscopy has been introduced in the context of strong-field induced pair creation \cite{JB2}.
Meanwhile, in the superposition of a nuclear Coulomb field and a high intensity laser field, we have demonstrated that the nucleus-induced deformation of the Dirac vacuum can be probed using phase-controlled colliding laser pulses \cite{CKL1}.

In this paper, we employ computational quantum field theory to investigate the creation of electron-positron pairs from vacuum induced by two spatiotemporally structured electric fields.
We analyze the  number of created electrons, their energy spectrum and the transition probability, all of which exhibit a strong dependence on the relative phase between the two fields.
Additionally, by relating the relative phase to the multi-photon transition channels through time-dependent perturbation theory, we show that the relative phase can be used to identify the order of multi-photon transitions.

Our paper is organized as follows. We start with a brief survey of the theoretical framework in Sec.\ref{2} and introduce the model system for the two spatiotemporal inhomogeneous electric fields with a relative phase.
In Sec.\ref{3}, we investigate the control of electron-positron pair creation via the relative phase between two electric fields, with a detailed analysis of its impact on various multi-photon processes. 
Finally, Sec.\ref{4} concludes the work.

\section{Theoretical Description of Pair Creation}
\label{2}
\subsection{The Computational Quantum Field Theory}
\label{2a}
To study the electron-positron pair creation process, the single particle quantum mechanical description is not sufficient and the quantum field theory is necessary to understand the particle creation and annihilation. 
In the framework of computational quantum field theory \cite{TC1,CKL2,MJ3,CKL3}, all dynamical features of the pair creation process are provided by the electron-positron field operator $\hat{\Psi}(z,t)$, whose space-time evolution is governed by the Dirac equation $i\hbar \partial \hat{\Psi}(z,t)= H \hat{\Psi}(z,t)$.  
Here the usual Hamiltonian in one spatial dimension is given by
\begin{equation}
H=c{\sigma}_{1} p_{z}+{\sigma}_{3} c^{2}+V(z, t),
 \label{eq1}
\end{equation} 
where $\sigma_{1}$ and $ \sigma_{3}$ are the $2 \times 2$ Pauli matrices. 
The electric field is described by the scalar potential $V(z,t)$. 
In the following, atomic units ($\hbar=m_e=e=1~a.u.$, and $c = 137.036~a.u.$) are used unless otherwise stated.
In our model, the external field is spatially uniform except for being localized along the $z$-axis. This allows us to simplify the Dirac four-component spinor wave functions into two components \cite{TC1,CKL2,MJ3,CKL3}.  
The field operator can be expanded as
\begin{align}
\hat{\psi}(z, t)&=\sum_{p} \hat{b}_{p}(t)u_{p}(z)+\sum_{n} \hat{d}_{n}^{\dagger}(t)v_{n}(z)  \notag \\
&=\sum_{p} \hat{b}_{p}u_{p}(z,t)+\sum_{n} \hat{d}_{n}^{\dagger}v_{n}(z,t),
 \label{eq2}
\end{align}
where $\hat{b}_{p}$ and $ \hat{d}^{\dagger}_{n} $ are the particle annihilation and the antiparticle creation operators, respectively.
The subscripts $p$ and $n$ correspond to positive and negative energy states. 
The fermionic annihilation and creation operators satisfy the anticommutation relations $ [\hat{b}_{p},\hat{b}^{\dagger}_{p^{'}}]_{+}=\delta_{p,p^{'}}$, $\quad[\hat{d}_{n},\hat{d}^{\dagger}_{n^{'}}]_{+}=\delta_{n,n^{'}}$.
The field-free Hamiltonian $h_{0}=c\sigma_{1}p_{z}+m\sigma_{3}c^{2}$ at $ t=0$ and its energy eigenstates are $ u_{p}(z) $ ($E \geq m c^{2}$) and $v_{n}(z)$ ($E \leq-m c^{2}$). 
$u_{p}(z,t)$ and  $v_{n}(z,t)$ are respectively the time evolved eigenstates of $u_{p}(z)$ and $ v_{n}(z)$, and their evolution satisfy the Dirac equation $i\hbar \partial \hat{\Psi}(z,t)= H \hat{\Psi}(z,t)$. 
The time evolution of the fermion annihilation and creation operators are
\begin{subequations}
\begin{align}
\hat{b}_{p}(t)=&\sum_{p^{'}} \hat{b}_{p^{'}}\int d z u_{p}^{*}(z) u_{p^{'}}(z, t) \notag \\
&\qquad+\sum_{n^{'}} \hat{d}^{\dagger}_{n^{'}}\int d z u_{p}^{*}(z) v_{n^{'}}(z, t), \label{eq3a} \\
 \hat{d}^{\dagger}_{n}(t)= &\sum_{p^{'}} \hat{b}_{p^{'}}\int d z v_{n}^{*}(z) u_{p^{'}}(z, t)  \notag \\
& \qquad+\sum_{n^{'}} \hat{d}^{\dagger}_{n^{'}} \int dz v_{n}^{*}(z) v_{n^{'}}(z, t),  \label{eq3b}
\end{align}
\end{subequations}
respectively. 
The electronic portion of the field operator $\hat{\psi}_{e}(z, t)\equiv\sum_{p} \hat{b}_{p}(t) u_{p}(z)$. 
With this definition, operators representing various physical quantities can be calculated, $e.g.$, the average spatial density of the created electrons
\begin{align}
\rho(z,t) &= \langle\langle\mathrm{vac} \| \hat{\psi}^{(e)\dag}(z,t) \hat{\psi}^{(e)}(z,t) \| \mathrm{vac}\rangle\rangle
\notag\\
&=\sum_n|\sum_p\mathbf{U}_{p,n}(t) u_p(z)|^2,
\label{eq4}
\end{align}
the momentum distribution
\begin{align}
\rho(p,t) &= \langle\langle\mathrm{vac} \| \hat{b}^{\dag}_{p}(t) \hat{b}_{p}(t) \| \mathrm{vac}\rangle\rangle
\notag\\
&=\sum_{n}\left|\mathbf{U}_{p,n}(t)\right|^{2},
\label{eq5}
\end{align}
and the total number of the created particle
\begin{align}
N(t)&=\int dz \langle\langle\mathrm{vac}\| \hat{\psi}_{e}^{\dagger}(z, t) \hat{\psi}_{e}(z, t)\| \mathrm{vac}\rangle\rangle \notag\\
&=\sum_{p}\langle\langle\mathrm{vac}\|\hat{b}^{\dagger}_{p}(t)\hat{b}_{p}(t)\|\mathrm{vac}\rangle\rangle,
\notag\\
&=\sum_{p, n}\left|\mathbf{U}_{p,n}(t)\right|^{2},
\label{eq6}
\end{align}
where $\mathbf{U}_{p,n}(t)=\int \,dzu_p^*(z)U(t)u_n(z)$ is a time evolution matrix.
The state $\|\mathrm{vac}\rangle\rangle$ represents the initial vacuum state, with the double bars and brackets indicating its status within second-quantized quantum field theory \cite{TC1,CKL2,MJ3,CKL3}.
In this scenario, all external fields were simultaneously turned off at that specific time and the resulting time-dependent number of created particles represents the actual physical count of particles \cite{CKL1,CCG1}.
The time evolution operator $U(t)=\hat{T}\exp\left[-i \int_{0}^{t} h(t^{'})dt^{'}\right]$ evolves the initial negative state $u_n(z)$ according to the single-particle Dirac equation.
This indicates that, for computational purposes, it is only necessary to evolve the set of negative energy continuum states $u_n(z)$.
The total time evolution from $0$ to $t_{\text{max}}$ is divided into $N_{T}$ intervals of $\Delta t = t_{\text{max}}/N_{T}$, each with an order of $10^{-6}~a.u.$.
The time evolution operator in each time step can be written as

\begin{align}
U(t+\Delta t,t)&=\hat{T}exp[-i\int_{t}^{t+\Delta t}(h_{0}+V(x,t))dt]\notag\\
&=exp[-i\int_{t}^{t+\Delta t}\dfrac{V(x,t)}{2}dt]\notag\\
&\times exp[-i\int_{t}^{t+\Delta t} h_{0}dt]\notag\\
&\times exp[-i\int_{t}^{t+\Delta t}\dfrac{V(x,t)}{2}dt]+O(\Delta t^{3})\notag\\
&\simeq exp(\dfrac{-iV\Delta t}{2}) \times  exp(-ih_{0}\Delta t)\notag\\
&\times exp(\dfrac{-iV\Delta t}{2})+O(\Delta t^{3}).
\label{eq7}
\end{align}
This is the split operator technique based on the third-order algorithm \cite{TC1,ADB1,GRM1,MF1}. 
By applying the fast Fourier transform between spatial and momentum space, the time evolution operator is decomposed into $N_{T}$ consecutive steps.

\subsection{The External Field Model}
\label{2b}
In the following, we examine two space-time dependent electric fields.
To simulate the space-time profile of the two electric fields, we chose for the left and right scalar potentials the form
\begin{subequations}
\begin{align}
V_r(z,t)&=S(z-z_0-d/2) sin(\omega t ), \label{eq8a} \\
V_l(z,t)&=S(z-z_0+d/2) sin(\omega t +\phi), \label{eq8b}
\end{align}
\end{subequations}
where $\omega$ denotes electric field frequency and $\phi$ denotes raletive phase between two fields. 
The left and right field are positioned at $z_0 - d/2$ and $z_0 + d/2$, respectively, with $d$ representing the distance between their centers.
In order to approximately characterize the basic space features of each electric field, we chose its spatial shape as $S(Z)=\frac{V_0}{2}\left[1+tanh(Z/W_0)\right]$.  
The height of field is $V_0$ and the spatial ramp up and down distance is $W_0$.
We choose for our total electric field configuration the superposition $V(z,t)=V_r(z,t)+V_l(z,t)$.

\begin{figure}[htbp]
\includegraphics[width=0.5\textwidth]{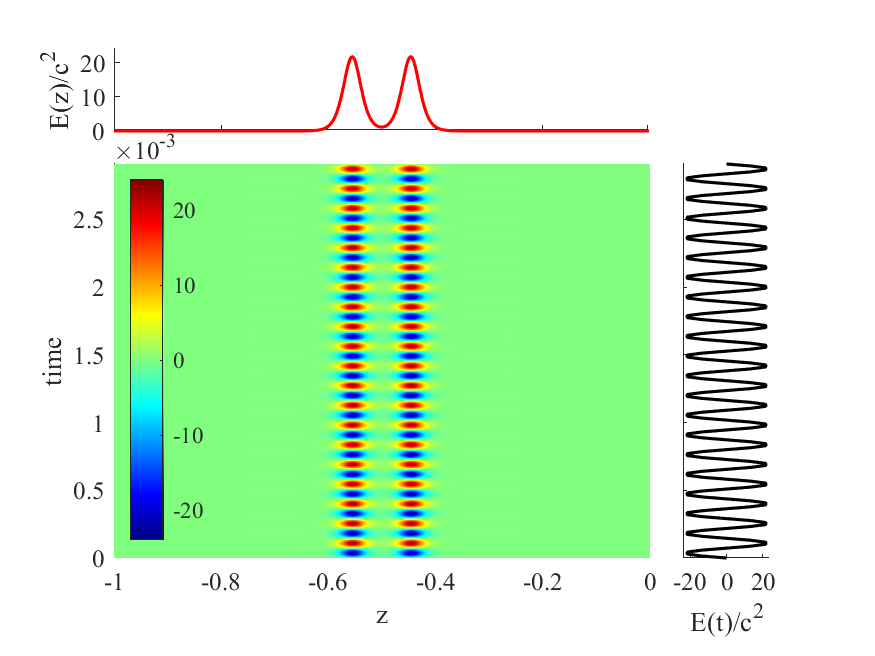}
\caption{Space-time resolved contour plot of the total electric profiled for $\phi=0$. The top panel shows the spatial profile along the red dashed line, while the right panel depicts the temporal oscillation of a single field along the white dashed line. The parameters are $V_0=c^2$, $W_0=3/c$, $z_0=L/2=0.5~a.u.$, $d=15/c$, $\omega=2.3c^2$, $t_{max}=20 \times 2\pi/\omega$.}
\label{fig1}
\end{figure}

As a consequence of our $1+1$ dimensional nature of our model quantum field theory, the transverse nature of a true electromagnetic field in three dimensions cannot be represented and there is no magnetic field.
The corresponding electric field is determined by taking the spatial derivative of the scalar potential: $E(z,t) = \partial V(z,t)/\partial z$.
To provide a clearer visualization of the total field, we have depicted its contour plot in Fig.\ref{fig1} within our numerical box of a total length $L=1~a.u.$.

\section{Results and Discussion}
\label{3}

\begin{figure}[htbp]
\includegraphics[width=0.5\textwidth]{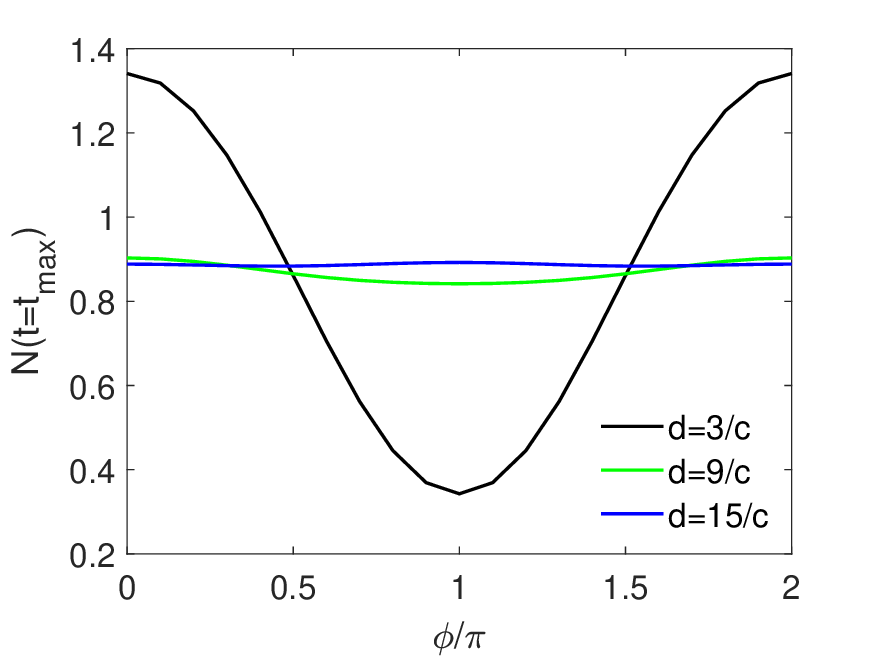}
\caption{The created electron-positron pairs number for different relative phase.
The black solid line, the green solid line and the blue solid line denotes distance $\d=3/c$, $9/c$ and $15/c$, respectively. Other parameters in our numerical simulations are the same as in Fig.~\ref{fig1}.}
\label{fig2}
\end{figure}

In this section, we investigate the control of electron-positron pair creation via the relative phase between two electric fields, with a detailed analysis of its impact on various multi-photon processes.
We first show how the number of created pairs depends on the relative phase.
To this end, we present numerically computed particle number in Fig.~\ref{fig2} for several values of the relative phase.
For distances $d = 3/c$ and $9/c$, the number of pairs first decreases and then increases with the relative phase, reaching a minimum at $\phi = \pi$, which resembles a cosine function. In contrast, for $d = 15/c$, the number of pairs remains constant and shows no dependence on the relative phase.
To understand this behavior, we examine the electric field profiles at different separations. We find that when the two field centers are brought close together, the tail of one scalar potential extends into the region of the other, causing the electric fields to overlap. In this regime, the resulting electric field amplitude is enhanced when the phases are matched and suppressed when they are opposite, which correspondingly enhances or suppresses the pair-creation process. Thus, at short distances, the number of created pairs can be effectively controlled by tuning the relative phase between the two fields.
For separations much larger than the spatial extent of the fields, the two electric fields exhibit negligible overlap and the number of created pairs remains unchanged.

\begin{figure}[htbp]
\includegraphics[width=0.5\textwidth]{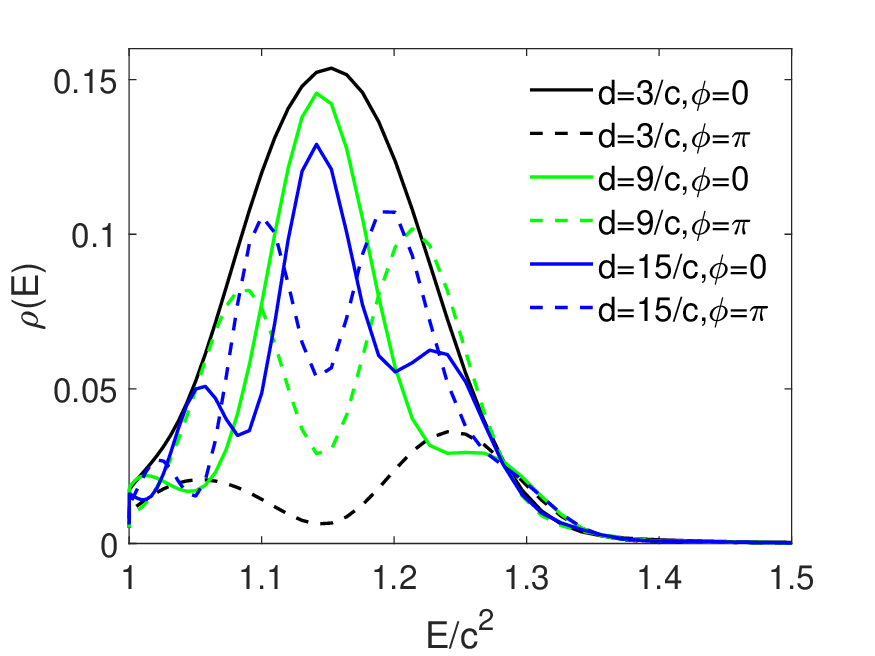}
\caption{The energy spectrum of the created particles.
The black line, the green line and the blue line denotes distance $d=3/c$, $9/c$ and $15/c$, respectively. The solid line represents the relative phase $\phi=0$, the dashed line represents the relative phase $\phi=\pi$. Other parameters in our numerical simulations are the same as in Fig.~\ref{fig1}.}
\label{fig3}
\end{figure}

In our model, the relative phases $\phi = 0$ and $\pi$ represent the two extreme cases at short distances. When the two fields are in phase ($\phi = 0$), the total strength of electric field is strongly enhanced and the number of created pairs reaches its maximum. By contrast, when they are in anti-phase ($\phi = \pi$), the total strength is strongly suppressed and the pair yield becomes minimal.
To illustrate this, we present in Fig.~\ref{fig3} the energy spectra of the created pairs for $\phi=0$ and $\pi$. At the distance $d=3/c$, the spectrum for $\phi=0$ is significantly enhanced, while that for $\phi=\pi$ is strongly suppressed, which is consistent with the results shown in Fig.~\ref{fig2}.
In particular, our findings reveal two distinct types of energy spectra for the created particles. For $\phi=0$, the relative phase locally modifies the energy spectrum while preserving the overall oscillatory structure dominated by the one-photon peak. Moreover, the number of peaks in the spectrum increases with distance. For $\phi=\pi$, the one-photon peak splits into two symmetric peaks, and its original location turns into a valley, which corresponds to multi-photon transition channels. These findings highlight the role of the relative phase in modulating the energy spectrum of the created particles.

\begin{figure*}[htbp]
\includegraphics[width=0.33\textwidth]{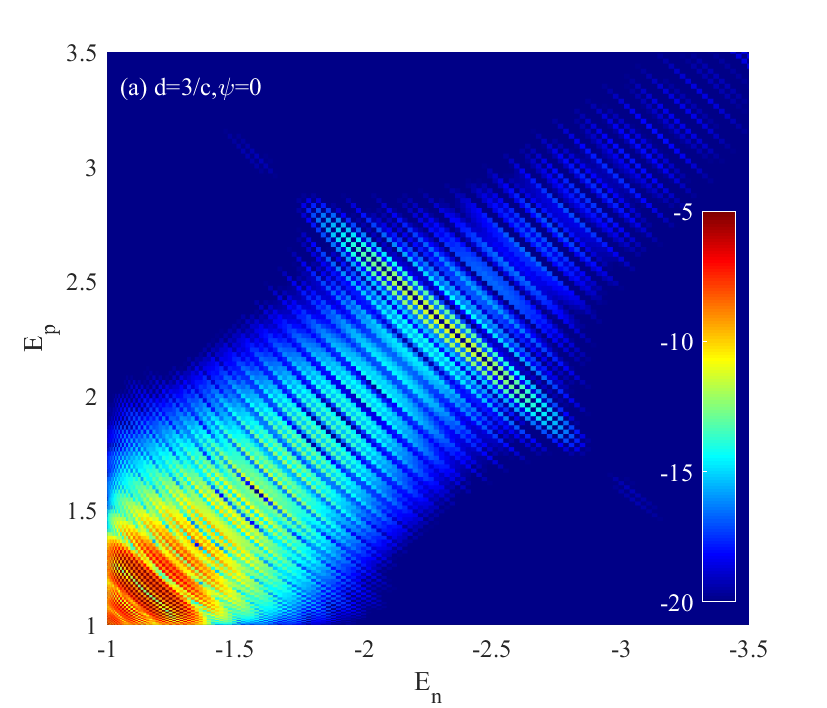}
\includegraphics[width=0.33\textwidth]{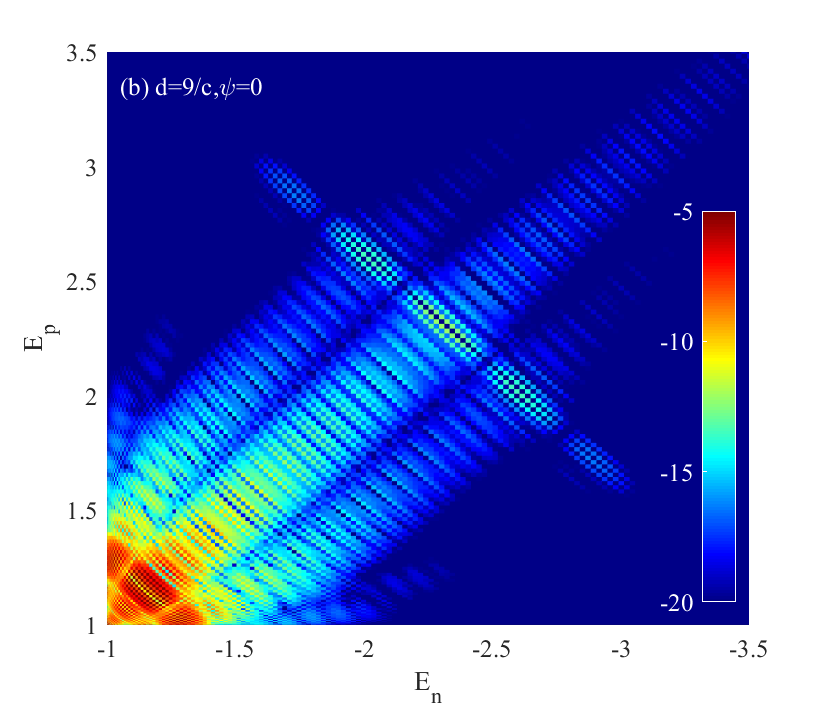}
\includegraphics[width=0.33\textwidth]{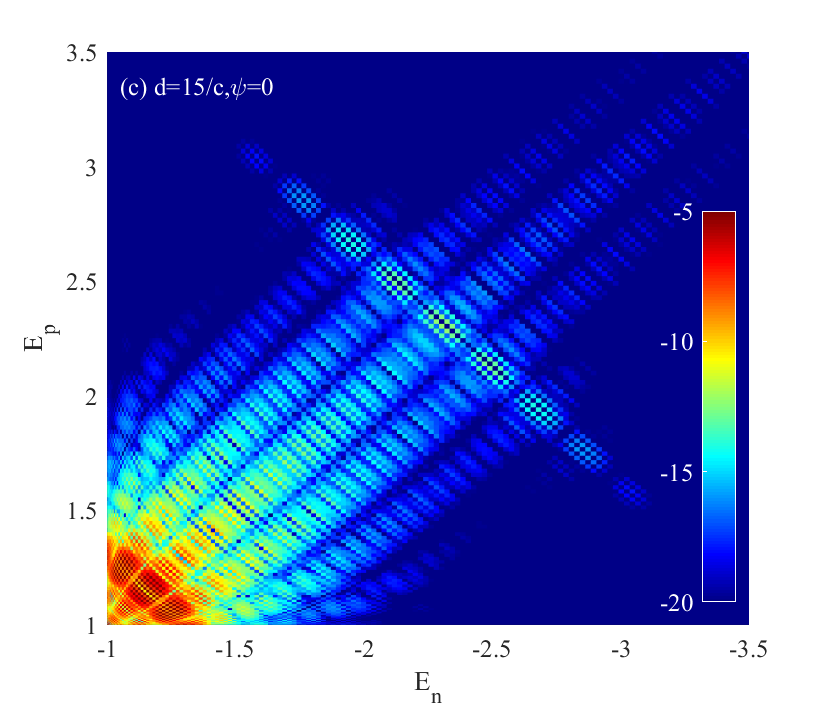}
\vspace{\baselineskip}
\includegraphics[width=0.33\textwidth]{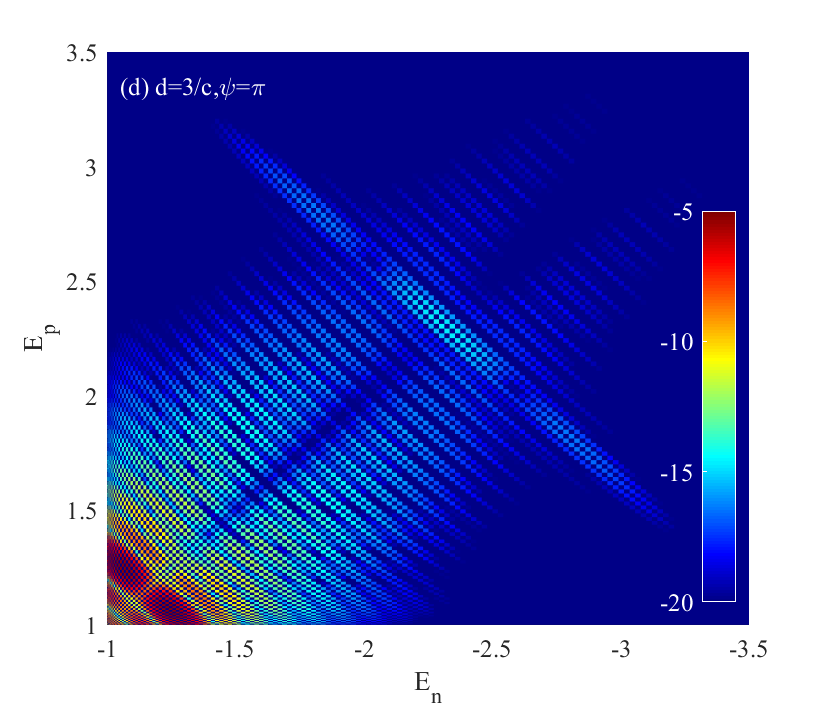}
\includegraphics[width=0.33\textwidth]{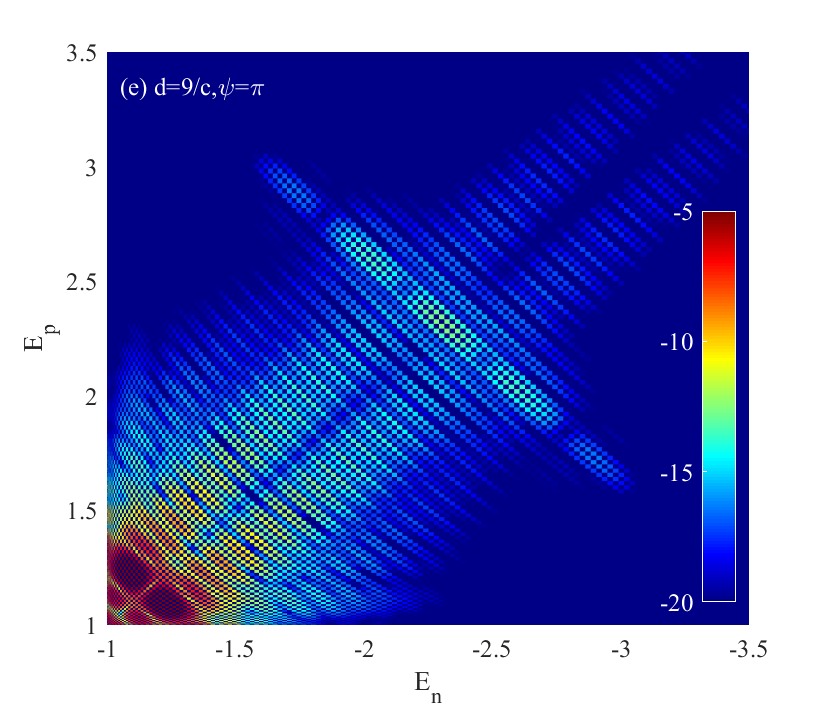}
\includegraphics[width=0.33\textwidth]{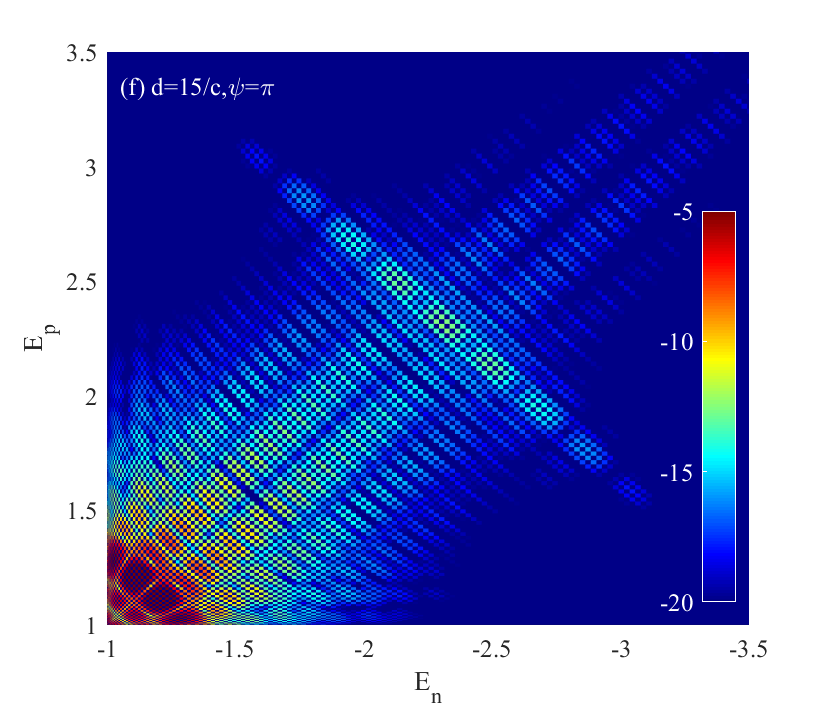}
\caption{The transition probability of electron for different relative phases. (a) $d=3/c,\phi=0$, (b) $d=9/c,\phi=0$ (c) $d=15/c,\phi=0$, (d) $d=3/c,\phi=\pi$, (e) $d=9/c, \phi=\pi$ and (f) $d=15/c, \phi=\pi$. The values on the color bar are presented on a logarithmic scale. Other parameters in our numerical simulations are the same as in Fig.~\ref{fig1}.}
\label{fig4}
\end{figure*}

In order to get a better idea how the relative phase influences multi-photon pair creation channels, we define the transition probability from a negative-energy state to a positive-energy state. 
The transition probability from the negative-energy state $|n\rangle$ to the positive-energy state $|p\rangle$ at time $t$ is given by $\left|\langle p | n(t) \rangle \right|^2$. 
In computational quantum field theory, this probability corresponds to the squared modulus of the time evolution matrix element, $\left| U_{p,n}(t) \right|^2$. 
The energy of the positive-energy state is $E_p=\sqrt{p^2c^2+c^4}$, while the energy of the negative-energy state is $E_n=-\sqrt{n^2c^2+c^4}$. 
Consequently, $\left| \langle E_p|E_n(t)\rangle \right|^{2}=\left|U_{E_p,E_n}(t)\right|^{2}$ can be interpreted as the probability of an electron transitioning from a negative-energy state $|E_n\rangle$ to a positive-energy state $|E_p\rangle$ at time $t$.
We illustrate the multi-photon transition probability between negative- and positive-energy states for the relative phases $\phi=0$ and $\phi=\pi$ in Fig.~\ref{fig4}. The transition probabilities are represented by the color scale, with the horizontal and vertical axes corresponding to negative and positive energy values, respectively. In accordance with energy conservation, one-photon transition channels predominantly occupy the lower energy regime, while two-photon channels are confined to the higher energy regime.

In Fig.~\ref{fig4}, two distinct channels for electron-positron pair creation are observed: the symmetric channel and the asymmetric channel. The diagonal line running from southwest to northeast corresponds to symmetric channels, which are characterized by symmetric energies satisfying $E_n = -E_p$. In contrast, the regions away from this diagonal represent asymmetric channels, where both the momenta and energies differ, with $E_n \neq -E_p$.
The competition between symmetric and asymmetric channels exhibits a pronounced dependence on the relative phase. The one-photon pathway differs for $\phi = \pi$ and $\phi = 0$, whereas the two-photon pathway remains unchanged. In particular, within the one-photon transition channel, symmetric transitions are forbidden when $\phi = \pi$ but allowed when $\phi = 0$. This demonstrates that multi-photon transition channels can be opened or closed by tuning the relative phase $\phi$.

\begin{figure*}[htbp]
\includegraphics[width=0.33\textwidth]{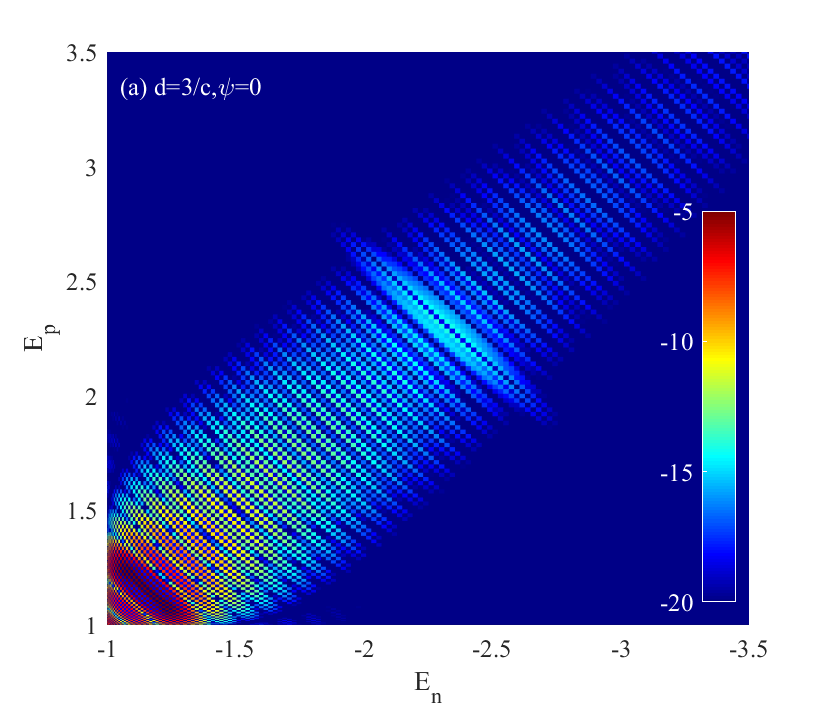}
\includegraphics[width=0.33\textwidth]{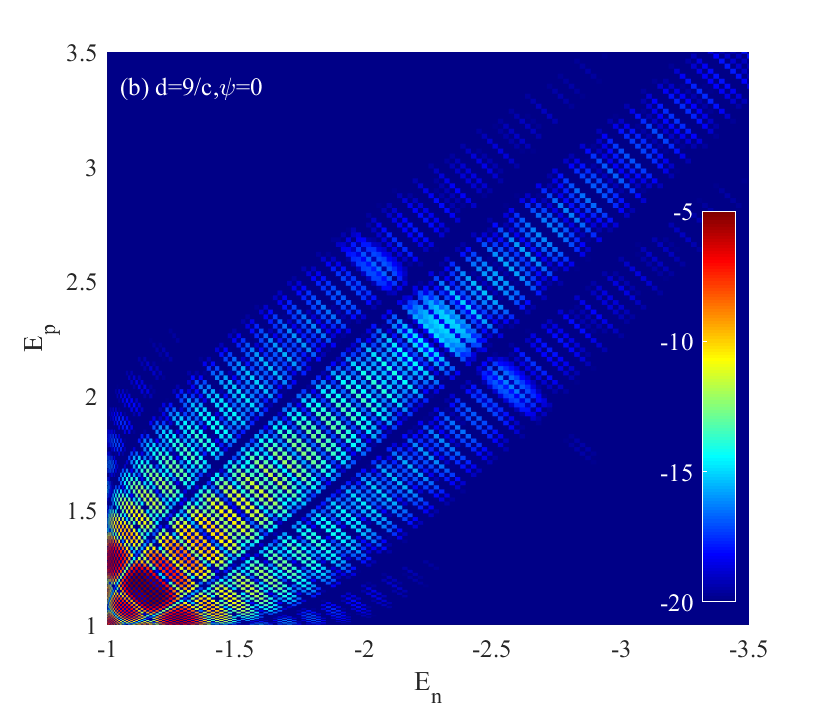}
\includegraphics[width=0.33\textwidth]{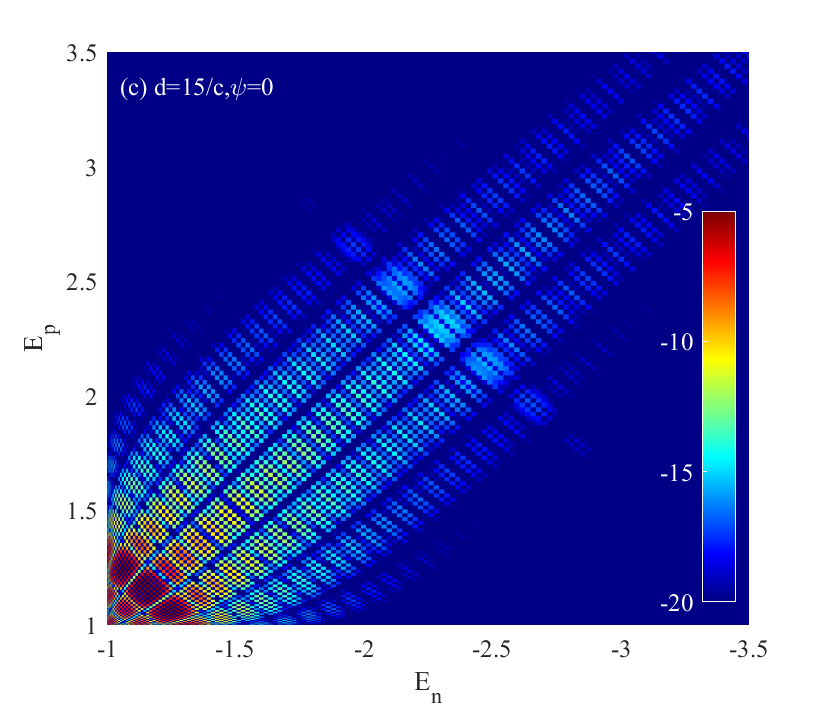}
\vspace{\baselineskip}
\includegraphics[width=0.33\textwidth]{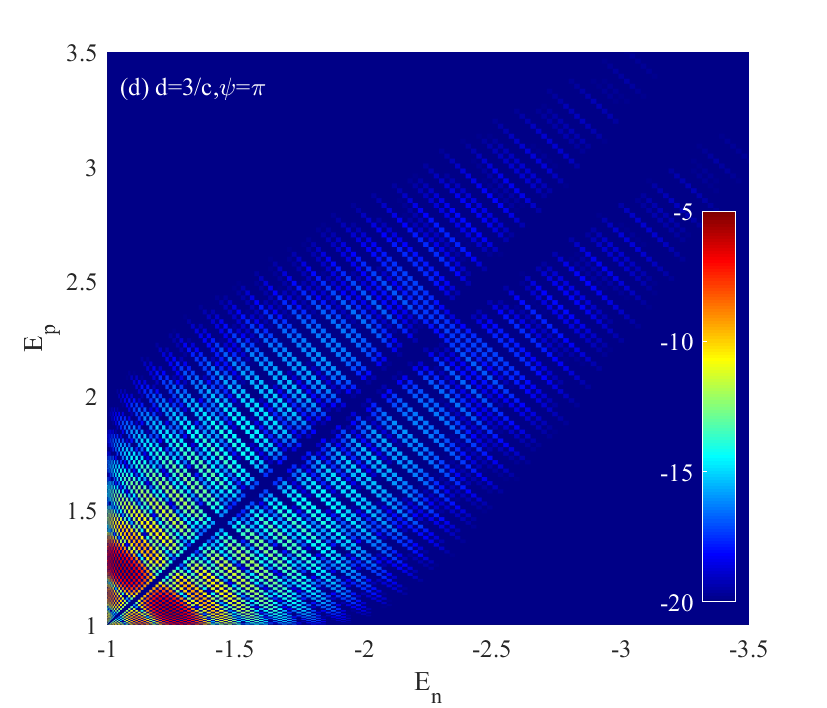}
\includegraphics[width=0.33\textwidth]{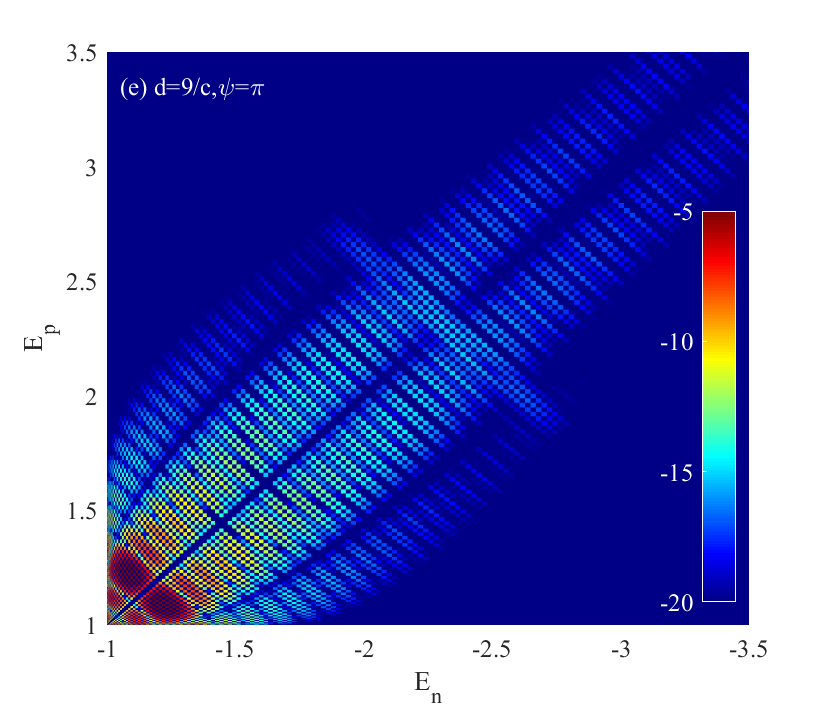}
\includegraphics[width=0.33\textwidth]{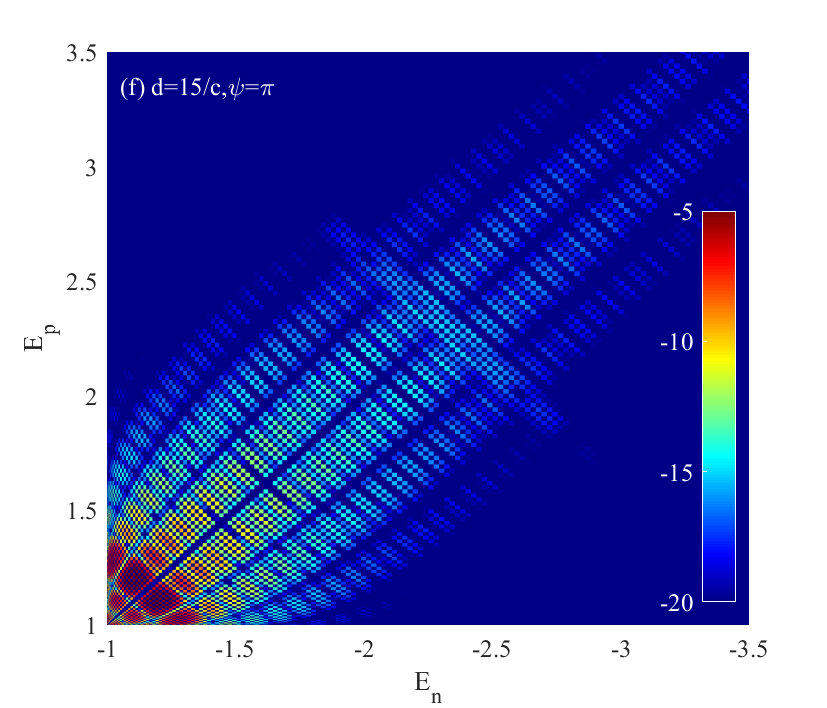}
\caption{The transition probability of electron computed by the perturbation theory. (a) $d=3/c,\phi=0$, (b) $d=9/c,\phi=0$ (c) $d=15/c,\phi=0$, (d) $d=3/c,\phi=\pi$, (e) $d=9/c, \phi=\pi$ and (f) $d=15/c, \phi=\pi$. The values on the color bar are presented on a logarithmic scale. Other parameters in our numerical simulations are the same as in Fig.~\ref{fig1}.}
\label{fig5}
\end{figure*}

In order to better understand this phenomenon, we apply the analytical framework outlined in Appendix~\ref{A}. Within time-dependent perturbation theory, the squared modulus of the transition amplitude $|C_{p,n}^{(1)}(t)\\+C_{p,n}^{(2)}(t)|^{2}$ gives the probability for a transition from the negative-energy state $|n\rangle$ to the positive-energy state $|p\rangle$ at time $t$. The electron transition probabilities predicted by time-dependent perturbation theory are illustrated in Fig.~\ref{fig5}. Excellent agreement is found with the results obtained from computational quantum field theory, confirming the validity of the perturbative description of multi-photon electron-positron pair creation. This amplitude thus serves as a bridge, establishing the theoretical consistency between computational quantum field theory and time-dependent perturbation theory in the present context.
From a theoretical perspective, the cosine term $\cos^2[(p+n)d/2+\phi/2]$ (see Eq.~\ref{eq10}) governs the symmetric transition pathways for the one-photon process. Symmetric transitions can occur only if the positive state $|p\rangle$ and the negative state $|n\rangle$ satisfy $p+n=0$. In this case, the cosine term $cos^2[(p+n)d/2+\phi/2]$ reduces to $cos^2(\phi/2)$. Consequently, symmetric transitions are open for $\phi=0$ but closed for $\phi=\pi$.
However, the two-photon process is considerably more complex (see Eq.~\ref{eq13}). For example, the exponential term $\exp[i(p+2k+n)d/2]$ indicates that this process involves intermediate states $|k\rangle$, adding further intricacy to the dynamics. These findings demonstrate the critical role of the relative phase in modulating multi-photon transition channels and reveal the distinct behaviors of one-photon and two-photon processes.

In addition, we observe that the interference pattern varies with the relative phase. As the distance $d$ increases, more interference fringes appear, which is consistent with the trend of the energy peaks in the energy spectrum (see Fig.~\ref{fig3}).
It is observed that different multi-photon transition processes lead to distinct interference patterns.
For the one-photon process, the interference pattern differs completely between $\phi=0$ and $\phi=\pi$. In contrast, two-photon transitions exhibit nearly identical interference characteristics for both phase configurations.
From the perspective of periodic variation with the relative phase, the one-photon process exhibits a period of $2\pi$, while the two-photon process has a period of $\pi$. Within first-order perturbation theory, the relative phase $\phi$ enters through the terms $\cos^2[(p+n)d/2+\phi/2]$ and $\lvert\exp(i\phi/2)\rvert^2$ in Eq.~\ref{eq10}. The exponential term equals $1$ and thus does not affect the energy spectrum of the created particles. However, the cosine term induces a periodic variation with a period of $2\pi$ as $\phi$ changes, giving rise to the observed interference effect. 
Similarly, in second-order perturbation theory, the terms $cos^2(\phi)$ and $\left|exp(i\phi)\right|^2$ in Eq.~\ref{eq13} are associated with the relative phase $\phi$. The exponential term also equals $1$, whereas the cosine term $\cos^2(\phi)$ introduces a periodic variation with a period of $\pi$. 
Notably, in perturbation theory, the first-order amplitude scales linearly with the perturbation, whereas the second-order amplitude scales quadratically. As a result, the phase dependence follows a simple trigonometric rule: for an $n$-photon process, the phase response period is expected to be $2\pi/n$. These findings establish the relative phase as an effective parameter for determining the order of multiphoton creation processes.

\section{Conclusion}
\label{4}
In summary, we have examined the electron-positron pair creation process induced by two oscillating electric fields with a relative phase $\phi$, utilizing computational quantum field theory.
We investigated the relationship between the relative phase and the number of created pairs. At small distances, the pair number varies with the relative phase in a cosine-like manner, while at large distances, it remains constant regardless of the relative phase.
We also examined the relationship between the relative phase and the energy spectra of the created particles, and we find that the relative phase locally modifies the energy spectrum. For the in-phase configuration ($\phi=0$), the overall oscillatory structure is preserved and remains dominated by a single one-photon peak. In the anti-phase case ($\phi=\pi$), by contrast, this peak splits into two peaks, with its original location turning into a valley.

The interference phenomena are observed in the electron transition probability during field-induced multi-photon pair creation. For one-photon processes, the transition probability varies periodically with the relative phase, exhibiting a period of $2\pi$. Notably, the transition probability of the two-photon processes remains unchanged when the relative phase is set to $\pi$.
Moreover, to provide a theoretical foundation for our findings, we employ a time-dependent perturbation theory framework. In this framework, the phase dependence arises naturally because the leading-order transition amplitudes are proportional to the perturbation strength and are governed by simple trigonometric relations. Within a simple perturbative treatment of an $n$-photon process, the transition probability is expected to exhibit a phase periodicity of $2\pi/n$.
It is demonstrated that the relative phase provides a reliable means to distinguish multi-photon transitions of different orders.

Furthermore, the relative phase can be used to control multi-photon transition channels. 
In first-order perturbation theory, the cosine term $\cos^2[(p+n)d/2+\phi/2]$ governs the symmetric transition pathways for the one-photon process. 
In particular, the opening and closing of symmetric one-photon transition channels can be precisely controlled by tuning the relative phase $\phi$. 
In second-order perturbation theory, the two-photon process is governed by a $\cos^2(\phi)$ dependence on the relative phase. 
However, the two-photon pair creation process is considerably more complex due to the involvement of intermediate states $|k\rangle$. 
These findings suggest that the relative phase can serve as an effective parameter for controlling electron-positron pair creation from vacuum.

\section*{Acknowledgements}
Fruitful discussions with D. D. Su, M. Jiang, Q. Chen and W. M. Wang are gratefully acknowledged. This work has been supported by the National Natural Science Foundation of China (NSFC) under Grants $Nos. 12\\447120$ and $12204001$ and Natural Science Foundation of Henan Province Grant $No. 252300423526$. Numerical simulations were implemented on the SongShan supercomputer at National Supercomputing Center in Zhengzhou.

\section*{Data Availability Statement}
This manuscript has no associated data. [Authors' comment: Data sharing not applicable to this article as no datasets were generated or analysed during the current study.]

\section*{Code Availability Statement}
This manuscript has no associated code/software. [Authors' comment: Code/Software sharing not applicable to this article as no code/software was generated or analysed during the current study.]

\appendix 
\renewcommand{\theequation}{\Alph{section}.\arabic{equation}} 
\section{The Time-dependent Perturbation Theory}
\label{A}
To understand the multi-photon transition mechanism, we recall that, for a space-time dependent electric field, the pair creation process can be viewed as a time-dependent perturbation problem. 
Since the interaction does not couple spin directions, we can express the field operator using only two components.
The Hamiltonian of this system can then be separated into two parts: $H=h_0+H^{'}$, where the external field is represented by $H^{'}=V(z,t)$, treated as a perturbation, and $h_0$ is the field-free Hamiltonian, analogous to the computational quantum field theory.
The first-order transition amplitude from the negative state $|n\rangle$ to the positive state $|p\rangle$ at an arbitrary time $t$ is given by 
\begin{equation}
C_{p,n}^{(1)}(t)=\int_{0}^{t}\langle p|H^{'}|n \rangle exp[i(E_p-E_n)\tau]/i~d\tau.
\label{eq9}
\end{equation}
Based on the rotating wave approximation, we obtain the following expression:
\begin{align}
C_{p,n}^{(1)}(t)&=-\pi W_0V_0A_{p,n}/4 csch[\pi W_0(p+n)/2]\notag\\
&\times \{exp[i(E_p-E_n-\omega)t]-1\} /(E_p-E_n-\omega)\notag\\
&\times cos[(p+n)d/2+\phi/2]exp(i\phi/2),
\label{eq10}
\end{align}
where the inner product $\langle p|n \rangle$ for spins $A_{p,n}$ is defined as $[sgn(n)\sqrt{E_p+c^2}\sqrt{-E_n-c^2}+sgn(p)\sqrt{E_p-c^2} \\
\sqrt{-E_n+c^2}]/[4\pi\sqrt{-E_pE_n}]$. 
The first-order perturbation estimate of created electrons $N_{P}^{(1)}(t)$ and its momentum distribution $\rho_{P}^{(1)}(p,t)$ is thus
\begin{equation}
N_{P}^{(1)}(t)=\sum_{p, n}\left|C_{p,n}^{(1)}(t)\right|^{2},
\label{eq11}
\end{equation}
\begin{equation}
\rho_{P}^{(1)}(p,t)=\sum_{n}\left|C_{p,n}^{(1)}(t)\right|^{2}.
\label{eq12}
\end{equation}

The time-dependent perturbation may be extended to higher orders as well, corresponding to multi-photon transitions, even though the magnitude due to these orders will be dramatically reduced. 
The amplitude of the second-order perturbation transition in time can be expressed as

\begin{flalign}
&C_{p,n}^{(2)}(t)=\pi^2 W_0^2V_0^2/32 \sum_{k}A_{p,k}csch[\pi W_0(p+k)/2] \notag\\
&\times A_{k,n}csch[\pi W_0(k+n)/2]\{exp[i(E_p-E_n-2\omega)t]-1\}\notag\\
&\times  1/(E_k-E_n-\omega)/(E_p-E_n-2\omega)\notag\\
&\times \{exp[i(p+2k+n)d/2]cos(\phi)+cos[(p+n)d/2]\}\notag\\
&\times exp(-i\phi),
\label{eq13}
\end{flalign}

where $A_{p,k}$ and $A_{k,n}$ can be obtained by the inner product $\langle p|k \rangle$ and $\langle k|n \rangle$.
Similarly, the number of created electrons $N_{P}^{(2)}(t)$ and corresponding momentum distribution $\rho_{P}^{(2)}(p,t)$ for the second-order perturbation is
\begin{equation}
N_{P}^{(2)}(t)=\sum_{p, n}\left|C_{p,n}^{(2)}(t)\right|^{2},
\label{eq14}
\end{equation}
\begin{equation}
\rho_{P}^{(2)}(p,t)=\sum_{n}\left|C_{p,n}^{(2)}(t)\right|^{2}.
\label{eq15}
\end{equation}
We obtain the final expression of the created pairs number and momentum distribution for the perturbation theory,
\begin{equation}
N_{P}(t)=\sum_{p, n}\left|C_{p,n}^{(1)}(t)+C_{p,n}^{(2)}(t)\right|^{2},
\label{eq16}
\end{equation}
\begin{equation}
\rho_{P}(p,t)=\sum_{n}\left|C_{p,n}^{(1)}(t)+C_{p,n}^{(2)}(t)\right|^{2}.
\label{eq17}
\end{equation}



\end{document}